\documentclass[aps,prb,showpacs,twocolumn,superscriptaddress]{revtex4-2}
\usepackage{appendix}
\usepackage{hyperref}
\usepackage{url}
\usepackage{amssymb}
\usepackage{amsmath}
\usepackage{graphicx}
\usepackage{epsfig}
\usepackage{subfigure}
\usepackage{amsfonts}
\usepackage{color}

\hypersetup{hypertex=true,
colorlinks=true,
linkcolor=blue,
urlcolor = blue,
citecolor=blue}

\begin{document}

\title{Detecting the Chern number via quench dynamics in two independent
chains}
\author{D. K. He}
\affiliation{School of Physics, Nankai University, Tianjin 300071, China}
\author{Y. B. Shi}
\email{1120210058@mail.nankai.edu.cn}
\affiliation{School of Physics, Nankai University, Tianjin 300071, China}
\author{Z. Song}
\email{songtc@nankai.edu.cn}
\affiliation{School of Physics, Nankai University, Tianjin 300071, China}

\begin{abstract}
The Chern number, as a topological invariant, characterizes the topological
features of a 2D system and can be experimentally detected through Hall
conductivity. In this work, we investigate the connection between the Chern
number and the features of two independent chains. It is shown that there
exists a class of 2D systems that can be mapped into two independent chains.
We demonstrate that the Chern number is identical to the linking number of
two loops, which are abstracted from each chain individually. This allows
for the detection of the Chern number via quench dynamics in two independent
chains. As an example, the Qi-Wu-Zhang (QWZ) model is employed to illustrate
the scheme. Our finding provides a way to measure the phase diagram of a 2D
system from the 1D systems.
\end{abstract}

\maketitle

\section{Introduction}

\label{Introduction} Topological phase, typically viewed as the equilibrium
properties characterized by topological invariants, goes beyond the Landau
paradigm and leads to topologically protected features, such as robustness
against local perturbations. Experiments on cold atoms enable the realization
of a wide variety of archetypal topological models accompanied by exotic
phenomena \cite%
{monroe2021programmable,blatt2012quantum,schreiber2015observation,bernien2017probing,choi2017observation,wallraff2004circuit,xu2020probing,chang2018colloquium,ye2008quantum,raimond2001manipulating,gring2012relaxation,neyenhuis2017observation,smith2016many,choi2016exploring,zhang2017observation1,zhang2017observation2,jurcevic2017direct}, including the Haldane model \cite{jotzu2014experimental,wu2016realization}, Thouless pumping \cite{lohse2016thouless}, the Harper--Hofstadter model 
\cite{tai2017microscopy} and its Bose-Einstein condensation \cite%
{kennedy2015observation}. A key advantage of studying topological systems on
such a platform is the ease of realizing dynamic processes \cite%
{flaschner2018observation}. Recent significant theoretical \cite%
{heyl2013dynamical,heyl2018dynamical,wang2017scheme} and experimental
advances \cite{jurcevic2017direct,zhang2017observation2} have greatly
enhanced our understanding of non-equilibrium behaviors \cite%
{zhang2017observation2}. This progress raises the natural question of
whether we can identify the topological boundary between two distinct phases
as revealed by the dynamical processes. One common approach is considering
the quench dynamics, i.e., preparing the non-interacting particles in the
the ground state of the pre-quenched Hamiltonian and driving them with the
post-quenched Hamiltonian. Several quantities, such as the rate function of
the Loschmidt echo \cite{heyl2015scaling} and the non-local order parameter
associated with the Bardeen-Cooper-Schrieffer (BCS)-like pairing channel 
\cite{shi2022dynamic} in Kitaev chain \cite%
{altland1997nonstandard,kitaev2001unpaired,soori2024majorana,soori2024majorana,decker2024density,malarddetecting,silva2024hybridization,starchl2022relaxation}, exhibit non-analytic behaviors when the quench crosses a phase boundary.
Such non-analytic behaviors can be easily unraveled based on the analogy
between quantum mechanics and classical electromagnetism \cite%
{shi2023emerging}. However, although most work has focused on one dimension,
the phenomena in higher dimensions have not been fully investigated \cite%
{kosior2024vortex}. It is highly desirable to investigate the
straightforward and experimentally feasible schemes to elucidate the
topological properties in\textbf{\ }high-dimensional system via quench
dynamics \cite{yang2020biot}.

In this work, we investigate the connection between the Chern number and the
features of two independent chains. It is shown that there exists a class of
2D systems that can be mapped into two independent chains. At equilibrium,
the bulk Chern number of the 2D system is known to correspond to the number
of edge states. This relationship is characterized by ``the bulk-edge
correspondence". However, there is no straightforward way to extract
topological information from the quench process and it is also difficult to
induce the intricate dynamics necessary for topological phases in the
high-dimensional systems. To address this issue, we investigate the linking
number of two loops abstracted from each chain individually and demonstrate
that the linking number is identical to the Chern number of the
corresponding 2D system. This result inspires us to investigate the quench
dynamics in two independent chains. The topological information can be
deduced from the performance of the evolved Bloch vectors. This approach
effectively reduces the system's dimensionality required for dynamics,
thereby simplifying the experimental complexity. We exemplify the
application of our approach based on the extended QWZ model \cite%
{qi2006topological}, which exhibits rich topological properties and serves
as the basic building block for the Bernevig-Hughes-Zhang model \cite%
{bernevig2006quantum} of the quantum spin Hall effect.

The remainder of this paper is organized as follows. In Section \ref{Model
and double loops}, we introduce the general form of the Bloch Hamiltonian
under investigation and demonstrate how to use the linking number of two
corresponding loops to predict its topological feature. In Section \ref%
{Hidden Chern number of two chains}, we present a representative
construction in the real space and demonstrate the approach to predict the
Chern number of a 2D model based on two 1D chains. In Section \ref%
{Nonequilibrium steady vector}, we propose a dynamical approach to detect
the topological information based on two independent quench processes driven
by the two chains respectively. Section \ref{Extended QWZ model} exemplifies
the application of our approach based on the extended QWZ model. Finally, we
present a summary and discussion in Section \ref{Summary and discussion}.

\begin{figure*}[tbh]
\centering\includegraphics[width=0.9\textwidth]{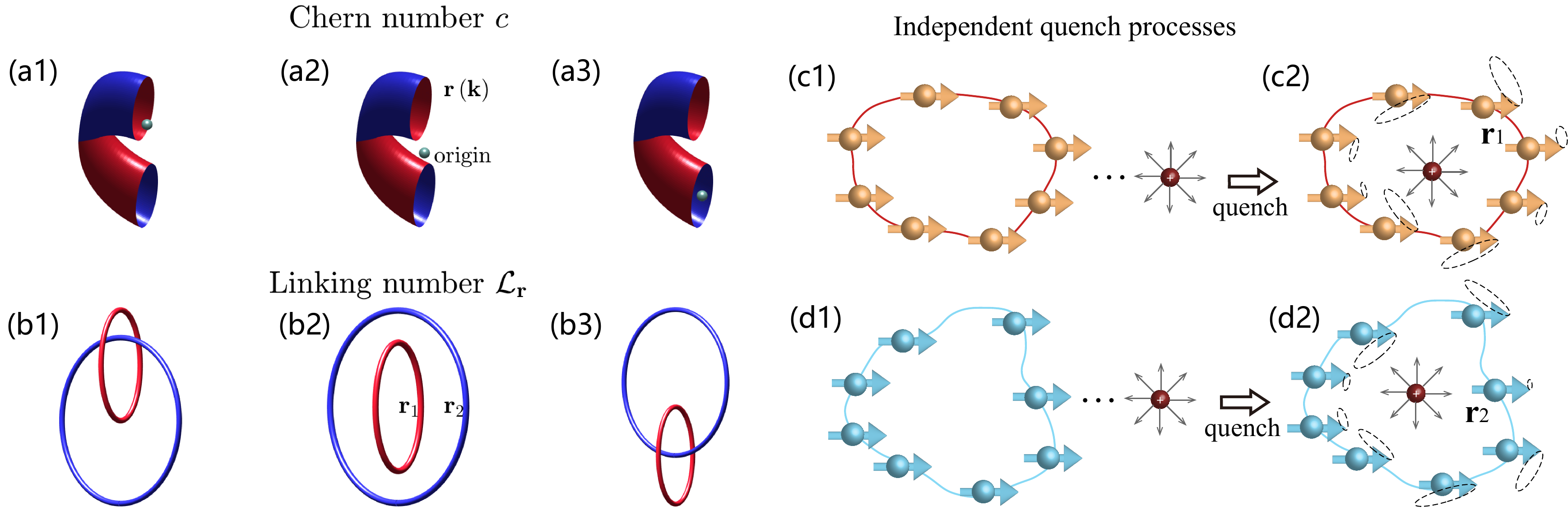}
\caption{(a1)-(a3) Schematic several representative configurations of the 
extended QWZ model. The topology of the system can be determined by
the times the torus $\mathbf{r\left(\mathbf{k}\right)}$ contains the origin
which is represented by the green solid circle. The blue and red
colors represent two sides of the surface. The potentials for panels (a1),
(a2), and (a3) are $\protect\mu =2,0,-3$, respectively, with the
corresponding Chern numbers $c = 1,0,-1$. For clarity, only the image of half
of the Brillouin zone is shown. Other parameters are $\protect\lambda_1 = 3$%
, $\protect\lambda_2 = 1$, $\protect\rho_1 =3$, and $\protect\rho_2 =2$.
(b1)-(b3) The Schematic representation of the two loops $\mathbf{r}_1$ and $%
\mathbf{r}_2$ defined in Eqs. (\protect\ref{QWZ_1}) and (\protect\ref{QWZ_2}%
). The two loops in (b) share the same parameters as the torus in (a) 
within the same column. The equivalence between the linking number $%
\mathcal{L}_{\mathbf{r}}$ and the Chern number $c$ enables the dynamical
detection. (c) and (d) Schematic illustration of the independent quench
processes. The red solid circle represents the origin, while the yellow and
cyan arrows indicate two sets of Bloch vectors. Initially, the Bloch vector
is prepared in the ground state of the prequench Hamiltonian and
subsequently driven by the postquench Hamiltonian. The topology of the 2D
model is determined by the behavior of the evolved Bloch vector. }
\label{fig1}
\end{figure*}

\section{Model and double loops}

\label{Model and double loops}

We begin by introducing the two-band Bloch Hamiltonian in the form $h\left( 
\mathbf{k}\right) =\mathbf{r}\left( \mathbf{k}\right) \mathbf{\cdot \sigma ,}
$where $\mathbf{\sigma =}\left( \sigma _{x},\sigma _{y},\sigma _{z}\right) $
are Pauli matrices satisfying the Lie algebra commutation relations, and
wave vector $\mathbf{k}=\left( k_{x},k_{y}\right) $ varies over the
Brillouin zone (BZ) $k_{x},k_{y}\in \left( -\pi ,\pi \right] $. The
eigenvalues of the Hamiltonian are $\pm \left\vert \mathbf{r}\left( \mathbf{k%
}\right) \right\vert $, corresponding to the upper and lower bands,
respectively, $h\left( \mathbf{k}\right) \left\vert u_{\pm }^{\mathbf{k}%
}\right\rangle =\pm \left\vert \mathbf{r}\left( \mathbf{k}\right)
\right\vert \left\vert u_{\pm }^{\mathbf{k}}\right\rangle $. We note that
the gap of $h\left( \mathbf{k}\right) $\ closes when the absolute value of $%
\mathbf{r}\left( \mathbf{k}\right) $\ equals zero. Matrix $h\left( \mathbf{k}%
\right) $ can serve as a core matrix of the crystalline system for
non-interacting Hamiltonian, or Kitaev chain. In equilibrium paradigm,
topological phase diagrams of this model can be characterized by Chern
number \cite{qi2011topological, cho2011quantum}, for which the lower band is
defined as,

\begin{equation}
c=\frac{1}{4\pi }\oint_{0}^{2\pi }\oint_{0}^{2\pi }\frac{\mathbf{r}}{%
\left\vert \mathbf{r}\right\vert ^{3}} \cdot \left( \partial _{k_{x}}\mathbf{r}_{1}%
\mathbf{\times }\partial _{k_{y}}\mathbf{r}_{2}\right) \mathrm{d}k_{x}%
\mathrm{d}k_{y}.  \label{CN}
\end{equation}%
It can also be interpreted as the number of times the surface traced by the
end of $\mathbf{r}\left( \mathbf{k}\right) $ in the auxiliary space encloses
the origin, as illustrated in Fig. \ref{fig1}(a).

In our paper, we focus on the condition that the vector $\mathbf{r}\left( 
\mathbf{k}\right) $\ can be represented by two periodic vector functions, $%
\mathbf{r}_{1}$ and $\mathbf{r}_{2}$, with respect to the wave vectors $%
k_{x},k_{y}$\ in the two directions, 
\begin{equation}
\mathbf{r}\left( k_{x},k_{y}\right) =\mathbf{r}_{1}\left( k_{x}\right) -%
\mathbf{r}_{2}(k_{y}).  \label{condition}
\end{equation}%
It is clear that \textbf{$\mathbf{r}_{1}$ }and\textbf{\ $\mathbf{r}_{2}$ }%
represent two loops in 3D auxiliary space. The gap closing $\left\vert 
\mathbf{r}\right\vert =0$ can be predicted by the crossing of them $\mathbf{r%
}_{1}\left( k_{x}\right) =\mathbf{r}_{2}(k_{y})$. This phenomenon inspires
us to explore the Chern number characterized by $\mathbf{r}\left( \mathbf{k}%
\right) $ based on the trajectories of two loops. To this end, we introduce
the corresponding linking number, 
\begin{equation}
\mathcal{L}_{\mathbf{r}}=\frac{1}{4\pi }\oint_{{\ell }_{2}}\oint_{{\ell }%
_{1}}\frac{\mathbf{r}_{1}-\mathbf{r}_{2}}{\left\vert \mathbf{r}_{1}-\mathbf{r%
}_{2}\right\vert ^{3}}  \cdot  \left( \mathrm{d}\mathbf{r}_{2}\times \mathrm{d}%
\mathbf{r}_{1}\right) .  \label{LN}
\end{equation}%
It can be readily verified that the linking number $\mathcal{L}_{\mathbf{r}}$
is equal to the Chern number $c$ defined in Eq. (\ref{CN}). This relation is
the main result in Ref. \cite{yang2020biot} and serves as the foundation of
our paper.

To describe this relation more concretely, we take the extended QWZ model as
a representative example, whose Bloch Hamiltonian is expressed as, 
\begin{equation}
\mathbf{r=}\left( \lambda _{x}\sin k_{x},\lambda _{y}\sin k_{y},\mu +\rho
_{x}\cos k_{x}+\rho _{y}\cos k_{y}\right) .  \label{QWZ}
\end{equation}%
This model was recently realized by the University of Science and Technology
of China and Peking University (USTC-PKU) group on a square Raman lattice 
\cite{yi2019observing}. In this setup, $\lambda $ and $\rho $ denote the
spin-conserved and spin-flip hopping coefficients, respectively \cite%
{ma2023orthogonal}. $\mu $ represents the Zeeman constant. Suddenly
modulating the Zeeman constants $\mu $ can implement the quench dynamics. We
can rewrite $\mathbf{r}$ in the form of $\mathbf{r}_{1}$ and $\mathbf{r}_{2}$%
, 
\begin{eqnarray}
\mathbf{r}_{1} &=&\left( \lambda _{x}\sin k_{x},0,\mu _{1}+\rho _{x}\cos
k_{x}\right) ,  \label{QWZ_1} \\
\mathbf{r}_{2} &=&\left( 0,-\lambda _{y}\sin k_{y},\mu _{2}-\rho _{y}\cos
k_{y}\right) ,  \label{QWZ_2}
\end{eqnarray}%
where $\mu _{1}-\mu _{2}=\mu $. Here, $\mathbf{r}_{1}$ and $\mathbf{r}_{2}$
represent two ovals in the $xz$ and $yz$ planes, respectively, with centers
at $\left( 0,0,\mu _{1}\right) $ and $\left( 0,0,\mu _{2}\right) $. We plot
several representative configurations in Fig. \ref{fig1}(a) and (b) for $%
\lambda _{x}=\rho _{x}=3$, $\lambda _{y}=1$, and $\rho _{y}=2$, while
varying the potential $\mu $ for the panels in different columns. The
schematics illustrate that the Chern number can be easily obtained from the
geometrical configuration of the two loops. Compared to the direct
calculation of the Chern number from the Berry connection, These relations
offer a convenient demonstration of the system's topological
characteristics. In the following, we present an intuitive application.

\begin{figure}[t]
\centering\includegraphics[width=0.46\textwidth]{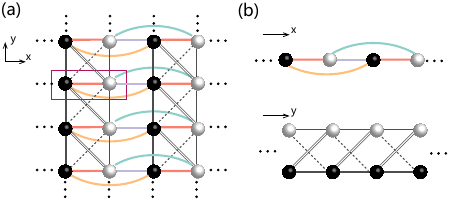}
\caption{(a) Schematic of the 2D model defined in Eq. (\protect\ref{2D}).
The unit cell consists of two sublattices, A (black) and B (white), as
indicated inside the red rectangle. The colored bonds represent the
interaction along the $x$ direction, while the black bonds, with varying
structures, indicate the interactions along the $y$ direction. (b) Schematic
of two independent chains corresponding to the 2D system in (a), as
demonstrated in Eqs. (\protect\ref{1D_1}) and (\protect\ref{1D_2}),
respectively. The colored bonds are used to construct one chain, and the
black bonds form the other. }
\label{fig2}
\end{figure}

\section{Hidden Chern number of two chains}

\label{Hidden Chern number of two chains}

Based on the above analysis, we attempt to reveal the topological properties
for a class of 2D systems using two 1D chains. These 2D systems comprise of
two sublattices, $A$ and $B$ with the coordinates for the unit cell denoted
as $(l,j)$, $1\leq l,j\leq N$, and can be realized in many experimental
platforms for the topological insulators \cite%
{mittal2018topological,hofmann2019chiral}. We use $a_{l,j}^{\dagger
},b_{l,j}^{\dagger }$ ($a_{l,j},b_{l,j}$) to represent the creation
(annihilation) operators for the sublattices $A$ and $B$. For the periodic
boundary condition considered in this paper, we set $a_{l+N,j+N}=a_{l,j}$
and $b_{l+N,j+N}=b_{l,j}$. The general schematic illustration is plotted in
Fig. \ref{fig2}(a). The coupling between the two unit cells occurs
exclusively along the $x$ or $y$ directions, represented by colored and
black bonds, respectively. In the $x$ direction, different colors indicate
varying coupling strengths, while the distinct structures of the black bonds
denote different coupling strengths in the $y$ direction. The general form
of the Hamiltonian in the real space is, 
\begin{equation}
H=\sum_{l,j,n}H_{l,j,n}^{x}+H_{l,j,n}^{y},  \label{2D}
\end{equation}%
where $H_{l,j,n}^{x}$ \textbf{(}$H_{l,j,n}^{y}$\textbf{)} denotes the $n$th
nearest neighbor coupling along $x$ ($y$)\ direction, 
\begin{eqnarray}
H_{l,j,n}^{x} &=&\left( a_{l,j}^{\dagger },b_{l,j}^{\dagger }\right) \mathbf{%
d}_{n}^{x}\cdot \mathbf{\sigma }\left( 
\begin{array}{c}
a_{l+n,j} \\ 
b_{l+n,j}%
\end{array}%
\right) +\mathrm{H.c.,} \\
H_{l,j,n}^{y} &=&\left( a_{l,j}^{\dagger },b_{l,j}^{\dagger }\right) \mathbf{%
d}_{n}^{y}\cdot \mathbf{\sigma }\left( 
\begin{array}{c}
a_{l,j+n} \\ 
b_{l,j+n}%
\end{array}%
\right) +\mathrm{H.c..}
\end{eqnarray}%
The vector components of $\mathbf{d}$\ can be real or complex. Applying the
Fourier transformation to the two sublattices, 
\begin{eqnarray}
a_{\mathbf{k}} &=&\frac{1}{N}\sum_{l,j}e^{-i\mathbf{k\cdot r}}a_{l,j}, \\
b_{\mathbf{k}} &=&\frac{1}{N}\sum_{l,j}e^{-i\mathbf{k\cdot r}}b_{l,j}.
\end{eqnarray}%
with the wave vector $\mathbf{k}=\left( k_{x},k_{y}\right) $\textbf{, }we
can rewrite the lattice Hamiltonian in the momentum space, 
\begin{equation}
H=\sum_{\mathbf{k}}\psi _{\mathbf{k}}^{\dagger }h\left( \mathbf{k}\right)
\psi _{\mathbf{k}}.
\end{equation}%
Here the basis is $\psi _{\mathbf{k}}=\left( a_{\mathbf{k}},b_{\mathbf{k}%
}\right) ^{T}$. The Bloch Hamiltonian is a $2\times 2$ matrix $h\left( 
\mathbf{k}\right) =\mathbf{r}\left( \mathbf{k}\right) \mathbf{\cdot \sigma }$
and the vector $\mathbf{r}\left( \mathbf{k}\right) $ takes the form 
\begin{equation}
\mathbf{r}\left( \mathbf{k}\right) =\sum_{n}\exp \left( ik_{x}n\right) 
\mathbf{d}_{n}^{x}+\exp \left( ik_{y}n\right) \mathbf{d}_{n}^{y}+\mathrm{c.c.%
}.  \label{RK}
\end{equation}%
Simultaneously, we can also construct two one-dimensional chains using the
coupling in the two directions, respectively, e.g., 
\begin{eqnarray}
H_{1} &=&\sum_{l,n}\left( a_{l,1}^{\dagger },b_{l,1}^{\dagger }\right) 
\mathbf{d}_{n}^{x}\cdot \mathbf{\sigma }\left( 
\begin{array}{c}
a_{l+n,1} \\ 
b_{l+n,1}%
\end{array}%
\right) +\mathrm{H.c.},  \label{1D_1} \\
H_{2} &=&-\sum_{j,n}\left( a_{1,j}^{\dagger },b_{1,j}^{\dagger }\right) 
\mathbf{d}_{n}^{y}\cdot \mathbf{\sigma }\left( 
\begin{array}{c}
a_{1,j+n} \\ 
b_{1,j+n}%
\end{array}%
\right) +\mathrm{H.c.,},  \label{1D_2}
\end{eqnarray}%
as illustrated in Fig. \ref{fig2}(b). The corresponding Bloch Hamiltonian $%
h_{\alpha }(k)=\mathbf{r}_{\alpha }(k)\mathbf{\cdot \sigma }$, $\alpha =1,2$
can be obtained easily based on the Fourier transformation in 1D and it is
easy to verify that $\mathbf{r\left( \mathbf{k}\right) =r}_{1}(k_{x})-%
\mathbf{r}_{2}(k_{y})$.

Based on the result in the previous section, we can directly conclude that
the topological phase of the 2D model in Eq. (\ref{2D}) can be completely
classified according to the linking number of two loops, $\mathbf{r}_{1}$
and $\mathbf{r}_{2}$, which are abstracted from each chain individually.
This description of the topological system is more feasible experimentally,
as it eliminates the need to fabricate a 2D topological structure, allowing
us to focus solely on two 1D chains.

\section{Nonequilibrium steady vector}

\label{Nonequilibrium steady vector}

In this section, we introduce a scheme to predict the topological properties
of 2D models via two independent quench processes driven by the 1D chains,
which is the primary focus of our work. An extensively studied approach to
analogize the quantum phase transition is known as dynamical quantum phase
transition (DQPT) \cite{heyl2013dynamical,heyl2018dynamical}. However, the
exact nature of the non-analyticity in 2D remains unclear currently.
Although several schemes \cite{shi2023emerging,wang2017scheme} are shown to
be applicable in 2D, it is worth investigating whether the topological
properties of the 2D model $H$ in Eq. (\ref{2D}) can be effectively captured
by the quench processes driven by $H_{1}$\ and $H_{2}$, given the challenges
of inducing the complex dynamics required for topological phases in
high-dimensional systems.

We thus consider the following quench process: Initially, we prepare two
sets of tensor product states, $\left\vert \psi _{\alpha }(0)\right\rangle
=\prod_{k}\left\vert \psi _{\alpha }^{k}(0)\right\rangle ,\alpha =1,2$. $%
\left\vert \psi _{\alpha }^{k}(0)\right\rangle $\ represents the eigenstate
of $\sigma _{z}$\ with eigenvalue $-1$. These two sets of initial states
contain no information about the driven Hamiltonian. Next, we drive them by $%
H_{\alpha }$ for any given time $t$ and every state will evolve as follows, 
\begin{equation}
\left\vert \psi _{\alpha }(t)\right\rangle =\exp \left( -iH_{\alpha
}t\right) \left\vert \psi _{\alpha }(0)\right\rangle ,\alpha =1,2.
\end{equation}%
It is easy to check that the evolved state remains a tensor product state,
i.e., $\left\vert \psi _{\alpha }(t)\right\rangle =\prod_{k}\exp \left[
-ih_{\alpha }(k)t\right] \left\vert \psi _{\alpha }^{k}(0)\right\rangle $ 
\textbf{.} We can calculate the interaction-free Bloch vector, 
\begin{equation}
\mathbf{n}_{\alpha }(k,t)=\left\langle \psi _{a}^{k}(t)\right\vert \mathbf{%
\sigma }\left\vert \psi _{a}^{k}(t)\right\rangle ,
\end{equation}%
for each momentum space over time. The time evolution of the Bloch vector
can be observed as precession along the direction of $\mathbf{r}_{\alpha
}(k,t)$ \cite{shi2023emerging}. A schematic illustration of the quench
process is plotted in Fig. \ref{fig1}(c) and (d). Initially, the endpoints
of Bloch vectors $\mathbf{n}_{\alpha }(k,t)$\ converge at the south pole on
the Bloch sphere, i.e., $\mathbf{n}_{1}(0)=\mathbf{n}_{2}(0)=\left(
0,0,-1\right) $. As time progresses, these endpoints will diverge and trace
out two instantaneous loops on the Bloch sphere. The dynamical behavior
allows us to calculate the following three variables: (1) The frequency of
precession, denoted as $\omega _{\alpha }^{k}$; (2) The average Bloch
vector, given by the formula 
\begin{equation}
\mathbf{\bar{n}}_{\alpha }(k,T)=\frac{1}{T}\int_{0}^{T}\mathbf{n}_{\alpha
}(k,\tau )\mathrm{d}\tau \text{;}  \label{In}
\end{equation}%
(3) The sign $\rho $ of the precession rate, where positive values indicate
a counterclockwise direction and negative values to indicate a clockwise
direction, as observed from the center of precession. We can delineate two
time-dependent loops based on the above three variables, 
\begin{equation}
\mathbf{l}_{\alpha }\left( k,T\right) =\frac{\rho \omega _{\alpha }^{k}%
\mathbf{\bar{n}}_{\alpha }(k,T)}{2\left\vert \mathbf{\bar{n}}_{\alpha
}(k,T)\right\vert },  \label{l}
\end{equation}%
and calculate the corresponding linking number 
\begin{equation}
\mathcal{L}_{\mathbf{l}}(T)=\frac{1}{4\pi }\oint \oint \frac{\mathbf{l}%
_{1}(T)-\mathbf{l}_{2}(T)}{\left\vert \mathbf{l}_{1}(T)-\mathbf{l}%
_{2}(T)\right\vert ^{3}}\cdot \left[ \mathrm{d}\mathbf{l}_{2}(T)\times 
\mathrm{d}\mathbf{l}_{1}(T)\right] .  \label{LN_T}
\end{equation}%
On the one hand, over an extended period, $T\rightarrow \infty $, the
integration in Eq. (\ref{In}) yields a steady value, given by, 
\begin{eqnarray}
\mathbf{\bar{n}}_{\alpha }(k) &=&\lim_{T\rightarrow \infty }\mathbf{\bar{n}}%
_{\alpha }(k,T) \\
&=&-\mathbf{\hat{r}}_{\alpha }\left( k\right) \left( \mathbf{\hat{r}}%
_{\alpha }\left( k\right) \cdot \widehat{z}\right) .
\end{eqnarray}%
The vector $\mathbf{\hat{r}}_{\alpha }\left( k\right) $ denotes the
direction of the Bloch Hamiltonian $\mathbf{r}_{\alpha }(k)$, $\mathbf{\hat{r%
}}_{\alpha }\left( k\right) =\mathbf{r}_{\alpha }\left( k\right) /\left\vert 
\mathbf{r}_{\alpha }\left( k\right) \right\vert $. On the other hand, it is
also straightforward to demonstrate that the precession frequency $\omega
_{\alpha }^{k}$ is equal to the bandwidth $2\left\vert \mathbf{r}_{\alpha
}\left( k\right) \right\vert $. Based on the above analysis, we can verify
that the trajectory of $\mathbf{l}_{\alpha }\left( k,T\right) $ will become
steady as time approaches infinity and converges to\textbf{\ $\mathbf{r}%
_{\alpha }\left( k\right) $.} The non-equilibrium steady behavior of $%
\mathbf{l}_{\alpha }\left( k,T\right) $ leads to the equivalence of the
linking number defined in Eq. (\ref{LN}) and Eq. (\ref{LN_T}),\ such that $%
\lim_{T\rightarrow \infty }\mathcal{L}_{\mathbf{l}}(T)=\mathcal{L}_{\mathbf{r%
}}$. In the following, we perform the numerical simulation based on the
extended QWZ model. 
\begin{figure}[t]
\centering\includegraphics[width=0.45\textwidth]{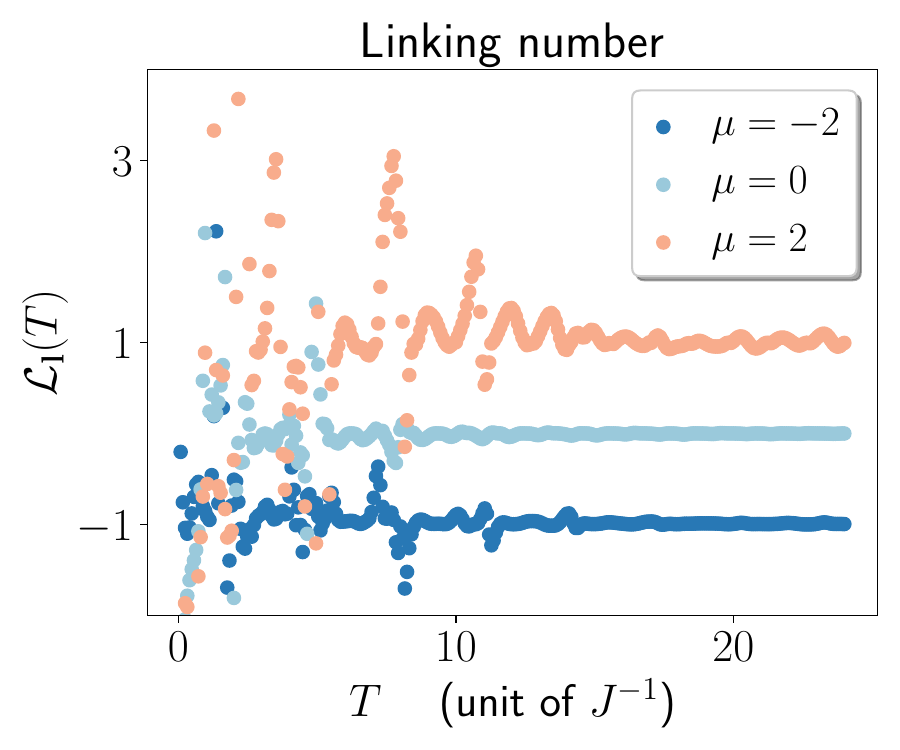}
\caption{Numerical simulations of the linking number defined in Eq. (\protect
\ref{LN_T}) as a function of time $T$ for case $N=50$ in various topological
regions. We can see that the linking number undergoes damped oscillation
around a stable value that matches the Chern number. Other parameters are $%
\protect\lambda _{x}=\protect\rho _{x}=3$, $\protect\lambda _{y}=1$, and $%
\protect\rho _{y}=2$. }
\label{fig3}
\end{figure}

\section{Extended QWZ model}

\label{Extended QWZ model} In this section, we consider the extended QWZ
model which has been introduced in Sec. \ref{Model and double loops} to
illustrate our scheme. To simplify the discussion, we select the same
parameters in Fig. \ref{fig1}(a) and (b) and the phase diagram is clear: $%
c=0$, for $\left\vert \mu \right\vert <1$ or $\left\vert \mu \right\vert >5$%
, and $c=\pm 1$ for $1<\pm \mu <5$.

We perform numerical simulation of the quench processes driven by the Bloch
Hamiltonian in Eqs. (\ref{QWZ_1}) and (\ref{QWZ_2}) for $\mu _{1}=\mu $ and $%
\mu _{2}=0$, and calculate $\mathcal{L}_{\mathbf{l}}(T)$ the as a function
of $T$. Here, the initial state $\left\vert \psi _{\alpha
}^{k}(0)\right\rangle =\left( 0,1\right) ^{T}$ can be prepared as the ground
state for infinite potential, $\mu _{1}\rightarrow \infty $ and $\mu
_{2}\rightarrow \infty $. It is accessible for the experiment in Ref. \cite%
{yi2019observing} as we can modulate the Zeeman constants flexibly. The
result is plotted in Fig. \ref{fig3}.

Initially, we find that the value of $\mathcal{L}_{\mathbf{l}}(T)$ is not an
integer during the early stages and oscillates irregularly. The reasons are
as follows: Referring to Eq. (\ref{l}), several wave vectors near the
critical value $k_{c}$ can lead to unpredictable and non-negligible errors
in numerical simulation. This critical value $k_{c}$ is defined by the
condition $\mathbf{\hat{r}}_{\alpha }\left( k_{c}\right) \cdot \widehat{z}=0$
and is also the source of non-analytic behavior in DQPT. Traditionally,
improving the accuracy of the simulation requires increasing the length $N$,
reducing the time interval $\delta t$ in dynamic simulations, adjusting data
types, and so on. These adjustments will inevitably lead to an increase in
simulation duration.

However, as time increases, the amplitudes of the oscillations continue to
decrease for various values of $\mu $ due to the integral in Eq. (\ref{In}). 
$\mathcal{L}_{\mathbf{l}}(T)$ tends to a steady value and eventually becomes
the same as the Chern number, thereby illustrating the topological
information of the QWZ model. The results indicate that our dynamical method
effectively conserves computational resources in numerical simulations.
Moreover, this approach may also facilitate experimental measurements by
offering a simplified and experimentally accessible way to evaluate the
topological features of 2D systems.

\section{Summary and discussion}

\label{Summary and discussion} In summary, we elucidate the topological
properties of 2D model $\mathbf{r}(k_{x},k_{y})$ through two 1D chains, $%
\mathbf{r}_{1}(k)$ and $\mathbf{r}_{2}(k)$, using a quantum quench dynamics
approach. we find that the equilibrium topological phase diagram of the 2D
model can be accurately predicted by the linking number of these
corresponding 1D chains. This equivalence inspires us to use a sudden quench
approach based on the two 1D chains to determine the Chern number of the 2D
model, which potentially facilitates experimental measurements. Based on the
exact solutions and numerical simulations, we observe that the steady
linking number of the two loops $\mathbf{l}_{1}\left( k,T\right) $ and $%
\mathbf{l}_{2}\left( k,T\right) $ generated by the quench dynamics displays
distinct behaviors depending on the topological phase of the system. The QWZ
model is employed as an example to illustrate the scheme. These findings
offer valuable insights into the dynamical detection of the topological
phases in 2D models using corresponding 1D chains. In future research, we
aim to extend our investigation to multi-band and three-dimensional models
and broaden our understanding of non-equilibrium behaviors.

\section*{Acknowledgment}

This work was supported by National Natural Science Foundation of China
(under Grant No. 12374461).


\begin{thebibliography}{46}%
	\makeatletter
	\providecommand \@ifxundefined [1]{%
		\@ifx{#1\undefined}
	}%
	\providecommand \@ifnum [1]{%
		\ifnum #1\expandafter \@firstoftwo
		\else \expandafter \@secondoftwo
		\fi
	}%
	\providecommand \@ifx [1]{%
		\ifx #1\expandafter \@firstoftwo
		\else \expandafter \@secondoftwo
		\fi
	}%
	\providecommand \natexlab [1]{#1}%
	\providecommand \enquote  [1]{``#1''}%
	\providecommand \bibnamefont  [1]{#1}%
	\providecommand \bibfnamefont [1]{#1}%
	\providecommand \citenamefont [1]{#1}%
	\providecommand \href@noop [0]{\@secondoftwo}%
	\providecommand \href [0]{\begingroup \@sanitize@url \@href}%
	\providecommand \@href[1]{\@@startlink{#1}\@@href}%
	\providecommand \@@href[1]{\endgroup#1\@@endlink}%
	\providecommand \@sanitize@url [0]{\catcode `\\12\catcode `\$12\catcode
		`\&12\catcode `\#12\catcode `\^12\catcode `\_12\catcode `\%12\relax}%
	\providecommand \@@startlink[1]{}%
	\providecommand \@@endlink[0]{}%
	\providecommand \url  [0]{\begingroup\@sanitize@url \@url }%
	\providecommand \@url [1]{\endgroup\@href {#1}{\urlprefix }}%
	\providecommand \urlprefix  [0]{URL }%
	\providecommand \Eprint [0]{\href }%
	\providecommand \doibase [0]{https://doi.org/}%
	\providecommand \selectlanguage [0]{\@gobble}%
	\providecommand \bibinfo  [0]{\@secondoftwo}%
	\providecommand \bibfield  [0]{\@secondoftwo}%
	\providecommand \translation [1]{[#1]}%
	\providecommand \BibitemOpen [0]{}%
	\providecommand \bibitemStop [0]{}%
	\providecommand \bibitemNoStop [0]{.\EOS\space}%
	\providecommand \EOS [0]{\spacefactor3000\relax}%
	\providecommand \BibitemShut  [1]{\csname bibitem#1\endcsname}%
	\let\auto@bib@innerbib\@empty
	\bibitem [{\citenamefont {Monroe}\ \emph {et~al.}(2021)\citenamefont {Monroe},
		\citenamefont {Campbell}, \citenamefont {Duan}, \citenamefont {Gong},
		\citenamefont {Gorshkov}, \citenamefont {Hess}, \citenamefont {Islam},
		\citenamefont {Kim}, \citenamefont {Linke}, \citenamefont {Pagano} \emph
		{et~al.}}]{monroe2021programmable}%
	\BibitemOpen
	\bibfield  {author} {\bibinfo {author} {\bibfnamefont {C.}~\bibnamefont
			{Monroe}}, \bibinfo {author} {\bibfnamefont {W.~C.}\ \bibnamefont
			{Campbell}}, \bibinfo {author} {\bibfnamefont {L.-M.}\ \bibnamefont {Duan}},
		\bibinfo {author} {\bibfnamefont {Z.-X.}\ \bibnamefont {Gong}}, \bibinfo
		{author} {\bibfnamefont {A.~V.}\ \bibnamefont {Gorshkov}}, \bibinfo {author}
		{\bibfnamefont {P.~W.}\ \bibnamefont {Hess}}, \bibinfo {author}
		{\bibfnamefont {R.}~\bibnamefont {Islam}}, \bibinfo {author} {\bibfnamefont
			{K.}~\bibnamefont {Kim}}, \bibinfo {author} {\bibfnamefont {N.~M.}\
			\bibnamefont {Linke}}, \bibinfo {author} {\bibfnamefont {G.}~\bibnamefont
			{Pagano}}, \emph {et~al.},\ }\href
	{https://doi.org/10.1103/RevModPhys.93.025001} {\bibfield  {journal}
		{\bibinfo  {journal} {Reviews of Modern Physics}\ }\textbf {\bibinfo {volume}
			{93}},\ \bibinfo {pages} {025001} (\bibinfo {year} {2021})}\BibitemShut
	{NoStop}%
	\bibitem [{\citenamefont {Blatt}\ and\ \citenamefont
		{Roos}(2012)}]{blatt2012quantum}%
	\BibitemOpen
	\bibfield  {author} {\bibinfo {author} {\bibfnamefont {R.}~\bibnamefont
			{Blatt}}\ and\ \bibinfo {author} {\bibfnamefont {C.~F.}\ \bibnamefont
			{Roos}},\ }\href {https://doi.org/10.1038/nphys2252} {\bibfield  {journal}
		{\bibinfo  {journal} {Nature Physics}\ }\textbf {\bibinfo {volume} {8}},\
		\bibinfo {pages} {277} (\bibinfo {year} {2012})}\BibitemShut {NoStop}%
	\bibitem [{\citenamefont {Schreiber}\ \emph {et~al.}(2015)\citenamefont
		{Schreiber}, \citenamefont {Hodgman}, \citenamefont {Bordia}, \citenamefont
		{L{\"u}schen}, \citenamefont {Fischer}, \citenamefont {Vosk}, \citenamefont
		{Altman}, \citenamefont {Schneider},\ and\ \citenamefont
		{Bloch}}]{schreiber2015observation}%
	\BibitemOpen
	\bibfield  {author} {\bibinfo {author} {\bibfnamefont {M.}~\bibnamefont
			{Schreiber}}, \bibinfo {author} {\bibfnamefont {S.~S.}\ \bibnamefont
			{Hodgman}}, \bibinfo {author} {\bibfnamefont {P.}~\bibnamefont {Bordia}},
		\bibinfo {author} {\bibfnamefont {H.~P.}\ \bibnamefont {L{\"u}schen}},
		\bibinfo {author} {\bibfnamefont {M.~H.}\ \bibnamefont {Fischer}}, \bibinfo
		{author} {\bibfnamefont {R.}~\bibnamefont {Vosk}}, \bibinfo {author}
		{\bibfnamefont {E.}~\bibnamefont {Altman}}, \bibinfo {author} {\bibfnamefont
			{U.}~\bibnamefont {Schneider}},\ and\ \bibinfo {author} {\bibfnamefont
			{I.}~\bibnamefont {Bloch}},\ }\href
	{https://www.science.org/doi/10.1126/science.aaa7432} {\bibfield  {journal}
		{\bibinfo  {journal} {Science}\ }\textbf {\bibinfo {volume} {349}},\ \bibinfo
		{pages} {842} (\bibinfo {year} {2015})}\BibitemShut {NoStop}%
	\bibitem [{\citenamefont {Bernien}\ \emph {et~al.}(2017)\citenamefont
		{Bernien}, \citenamefont {Schwartz}, \citenamefont {Keesling}, \citenamefont
		{Levine}, \citenamefont {Omran}, \citenamefont {Pichler}, \citenamefont
		{Choi}, \citenamefont {Zibrov}, \citenamefont {Endres}, \citenamefont
		{Greiner} \emph {et~al.}}]{bernien2017probing}%
	\BibitemOpen
	\bibfield  {author} {\bibinfo {author} {\bibfnamefont {H.}~\bibnamefont
			{Bernien}}, \bibinfo {author} {\bibfnamefont {S.}~\bibnamefont {Schwartz}},
		\bibinfo {author} {\bibfnamefont {A.}~\bibnamefont {Keesling}}, \bibinfo
		{author} {\bibfnamefont {H.}~\bibnamefont {Levine}}, \bibinfo {author}
		{\bibfnamefont {A.}~\bibnamefont {Omran}}, \bibinfo {author} {\bibfnamefont
			{H.}~\bibnamefont {Pichler}}, \bibinfo {author} {\bibfnamefont
			{S.}~\bibnamefont {Choi}}, \bibinfo {author} {\bibfnamefont {A.~S.}\
			\bibnamefont {Zibrov}}, \bibinfo {author} {\bibfnamefont {M.}~\bibnamefont
			{Endres}}, \bibinfo {author} {\bibfnamefont {M.}~\bibnamefont {Greiner}},
		\emph {et~al.},\ }\href {https://doi.org/10.1038/nature24622} {\bibfield
		{journal} {\bibinfo  {journal} {Nature}\ }\textbf {\bibinfo {volume} {551}},\
		\bibinfo {pages} {579} (\bibinfo {year} {2017})}\BibitemShut {NoStop}%
	\bibitem [{\citenamefont {Choi}\ \emph {et~al.}(2017)\citenamefont {Choi},
		\citenamefont {Choi}, \citenamefont {Landig}, \citenamefont {Kucsko},
		\citenamefont {Zhou}, \citenamefont {Isoya}, \citenamefont {Jelezko},
		\citenamefont {Onoda}, \citenamefont {Sumiya}, \citenamefont {Khemani} \emph
		{et~al.}}]{choi2017observation}%
	\BibitemOpen
	\bibfield  {author} {\bibinfo {author} {\bibfnamefont {S.}~\bibnamefont
			{Choi}}, \bibinfo {author} {\bibfnamefont {J.}~\bibnamefont {Choi}}, \bibinfo
		{author} {\bibfnamefont {R.}~\bibnamefont {Landig}}, \bibinfo {author}
		{\bibfnamefont {G.}~\bibnamefont {Kucsko}}, \bibinfo {author} {\bibfnamefont
			{H.}~\bibnamefont {Zhou}}, \bibinfo {author} {\bibfnamefont {J.}~\bibnamefont
			{Isoya}}, \bibinfo {author} {\bibfnamefont {F.}~\bibnamefont {Jelezko}},
		\bibinfo {author} {\bibfnamefont {S.}~\bibnamefont {Onoda}}, \bibinfo
		{author} {\bibfnamefont {H.}~\bibnamefont {Sumiya}}, \bibinfo {author}
		{\bibfnamefont {V.}~\bibnamefont {Khemani}}, \emph {et~al.},\ }\href
	{https://doi.org/10.1038/nature21426} {\bibfield  {journal} {\bibinfo
			{journal} {Nature}\ }\textbf {\bibinfo {volume} {543}},\ \bibinfo {pages}
		{221} (\bibinfo {year} {2017})}\BibitemShut {NoStop}%
	\bibitem [{\citenamefont {Wallraff}\ \emph {et~al.}(2004)\citenamefont
		{Wallraff}, \citenamefont {Schuster}, \citenamefont {Blais}, \citenamefont
		{Frunzio}, \citenamefont {Huang}, \citenamefont {Majer}, \citenamefont
		{Kumar}, \citenamefont {Girvin},\ and\ \citenamefont
		{Schoelkopf}}]{wallraff2004circuit}%
	\BibitemOpen
	\bibfield  {author} {\bibinfo {author} {\bibfnamefont {A.}~\bibnamefont
			{Wallraff}}, \bibinfo {author} {\bibfnamefont {D.}~\bibnamefont {Schuster}},
		\bibinfo {author} {\bibfnamefont {A.}~\bibnamefont {Blais}}, \bibinfo
		{author} {\bibfnamefont {L.}~\bibnamefont {Frunzio}}, \bibinfo {author}
		{\bibfnamefont {R.-S.}\ \bibnamefont {Huang}}, \bibinfo {author}
		{\bibfnamefont {J.}~\bibnamefont {Majer}}, \bibinfo {author} {\bibfnamefont
			{S.}~\bibnamefont {Kumar}}, \bibinfo {author} {\bibfnamefont
			{S.}~\bibnamefont {Girvin}},\ and\ \bibinfo {author} {\bibfnamefont
			{R.}~\bibnamefont {Schoelkopf}},\ }\href
	{https://doi.org/10.48550/arXiv.cond-mat/0407325} {\bibfield  {journal}
		{\bibinfo  {journal} {arXiv preprint cond-mat/0407325}\ } (\bibinfo {year}
		{2004})}\BibitemShut {NoStop}%
	\bibitem [{\citenamefont {Xu}\ \emph {et~al.}(2020)\citenamefont {Xu},
		\citenamefont {Sun}, \citenamefont {Liu}, \citenamefont {Zhang},
		\citenamefont {Li}, \citenamefont {Dong}, \citenamefont {Ren}, \citenamefont
		{Zhang}, \citenamefont {Nori}, \citenamefont {Zheng} \emph
		{et~al.}}]{xu2020probing}%
	\BibitemOpen
	\bibfield  {author} {\bibinfo {author} {\bibfnamefont {K.}~\bibnamefont
			{Xu}}, \bibinfo {author} {\bibfnamefont {Z.-H.}\ \bibnamefont {Sun}},
		\bibinfo {author} {\bibfnamefont {W.}~\bibnamefont {Liu}}, \bibinfo {author}
		{\bibfnamefont {Y.-R.}\ \bibnamefont {Zhang}}, \bibinfo {author}
		{\bibfnamefont {H.}~\bibnamefont {Li}}, \bibinfo {author} {\bibfnamefont
			{H.}~\bibnamefont {Dong}}, \bibinfo {author} {\bibfnamefont {W.}~\bibnamefont
			{Ren}}, \bibinfo {author} {\bibfnamefont {P.}~\bibnamefont {Zhang}}, \bibinfo
		{author} {\bibfnamefont {F.}~\bibnamefont {Nori}}, \bibinfo {author}
		{\bibfnamefont {D.}~\bibnamefont {Zheng}}, \emph {et~al.},\ }\href
	{https://www.science.org/doi/10.1126/sciadv.aba4935} {\bibfield  {journal}
		{\bibinfo  {journal} {Science advances}\ }\textbf {\bibinfo {volume} {6}},\
		\bibinfo {pages} {eaba4935} (\bibinfo {year} {2020})}\BibitemShut {NoStop}%
	\bibitem [{\citenamefont {Chang}\ \emph {et~al.}(2018)\citenamefont {Chang},
		\citenamefont {Douglas}, \citenamefont {Gonz{\'a}lez-Tudela}, \citenamefont
		{Hung},\ and\ \citenamefont {Kimble}}]{chang2018colloquium}%
	\BibitemOpen
	\bibfield  {author} {\bibinfo {author} {\bibfnamefont {D.}~\bibnamefont
			{Chang}}, \bibinfo {author} {\bibfnamefont {J.}~\bibnamefont {Douglas}},
		\bibinfo {author} {\bibfnamefont {A.}~\bibnamefont {Gonz{\'a}lez-Tudela}},
		\bibinfo {author} {\bibfnamefont {C.-L.}\ \bibnamefont {Hung}},\ and\
		\bibinfo {author} {\bibfnamefont {H.}~\bibnamefont {Kimble}},\ }\href
	{https://doi.org/10.1103/RevModPhys.90.031002} {\bibfield  {journal}
		{\bibinfo  {journal} {Reviews of Modern Physics}\ }\textbf {\bibinfo {volume}
			{90}},\ \bibinfo {pages} {031002} (\bibinfo {year} {2018})}\BibitemShut
	{NoStop}%
	\bibitem [{\citenamefont {Ye}\ \emph {et~al.}(2008)\citenamefont {Ye},
		\citenamefont {Kimble},\ and\ \citenamefont {Katori}}]{ye2008quantum}%
	\BibitemOpen
	\bibfield  {author} {\bibinfo {author} {\bibfnamefont {J.}~\bibnamefont
			{Ye}}, \bibinfo {author} {\bibfnamefont {H.}~\bibnamefont {Kimble}},\ and\
		\bibinfo {author} {\bibfnamefont {H.}~\bibnamefont {Katori}},\ }\href
	{https://www.science.org/doi/10.1126/science.1148259} {\bibfield  {journal}
		{\bibinfo  {journal} {science}\ }\textbf {\bibinfo {volume} {320}},\ \bibinfo
		{pages} {1734} (\bibinfo {year} {2008})}\BibitemShut {NoStop}%
	\bibitem [{\citenamefont {Raimond}\ \emph {et~al.}(2001)\citenamefont
		{Raimond}, \citenamefont {Brune},\ and\ \citenamefont
		{Haroche}}]{raimond2001manipulating}%
	\BibitemOpen
	\bibfield  {author} {\bibinfo {author} {\bibfnamefont {J.-M.}\ \bibnamefont
			{Raimond}}, \bibinfo {author} {\bibfnamefont {M.}~\bibnamefont {Brune}},\
		and\ \bibinfo {author} {\bibfnamefont {S.}~\bibnamefont {Haroche}},\ }\href
	{https://doi.org/10.1103/RevModPhys.73.565} {\bibfield  {journal} {\bibinfo
			{journal} {Reviews of Modern Physics}\ }\textbf {\bibinfo {volume} {73}},\
		\bibinfo {pages} {565} (\bibinfo {year} {2001})}\BibitemShut {NoStop}%
	\bibitem [{\citenamefont {Gring}\ \emph {et~al.}(2012)\citenamefont {Gring},
		\citenamefont {Kuhnert}, \citenamefont {Langen}, \citenamefont {Kitagawa},
		\citenamefont {Rauer}, \citenamefont {Schreitl}, \citenamefont {Mazets},
		\citenamefont {Smith}, \citenamefont {Demler},\ and\ \citenamefont
		{Schmiedmayer}}]{gring2012relaxation}%
	\BibitemOpen
	\bibfield  {author} {\bibinfo {author} {\bibfnamefont {M.}~\bibnamefont
			{Gring}}, \bibinfo {author} {\bibfnamefont {M.}~\bibnamefont {Kuhnert}},
		\bibinfo {author} {\bibfnamefont {T.}~\bibnamefont {Langen}}, \bibinfo
		{author} {\bibfnamefont {T.}~\bibnamefont {Kitagawa}}, \bibinfo {author}
		{\bibfnamefont {B.}~\bibnamefont {Rauer}}, \bibinfo {author} {\bibfnamefont
			{M.}~\bibnamefont {Schreitl}}, \bibinfo {author} {\bibfnamefont
			{I.}~\bibnamefont {Mazets}}, \bibinfo {author} {\bibfnamefont {D.~A.}\
			\bibnamefont {Smith}}, \bibinfo {author} {\bibfnamefont {E.}~\bibnamefont
			{Demler}},\ and\ \bibinfo {author} {\bibfnamefont {J.}~\bibnamefont
			{Schmiedmayer}},\ }\href {https://doi.org/10.1126/science.1224953} {\bibfield
		{journal} {\bibinfo  {journal} {Science}\ }\textbf {\bibinfo {volume}
			{337}},\ \bibinfo {pages} {1318} (\bibinfo {year} {2012})}\BibitemShut
	{NoStop}%
	\bibitem [{\citenamefont {Neyenhuis}\ \emph {et~al.}(2017)\citenamefont
		{Neyenhuis}, \citenamefont {Zhang}, \citenamefont {Hess}, \citenamefont
		{Smith}, \citenamefont {Lee}, \citenamefont {Richerme}, \citenamefont {Gong},
		\citenamefont {Gorshkov},\ and\ \citenamefont
		{Monroe}}]{neyenhuis2017observation}%
	\BibitemOpen
	\bibfield  {author} {\bibinfo {author} {\bibfnamefont {B.}~\bibnamefont
			{Neyenhuis}}, \bibinfo {author} {\bibfnamefont {J.}~\bibnamefont {Zhang}},
		\bibinfo {author} {\bibfnamefont {P.~W.}\ \bibnamefont {Hess}}, \bibinfo
		{author} {\bibfnamefont {J.}~\bibnamefont {Smith}}, \bibinfo {author}
		{\bibfnamefont {A.~C.}\ \bibnamefont {Lee}}, \bibinfo {author} {\bibfnamefont
			{P.}~\bibnamefont {Richerme}}, \bibinfo {author} {\bibfnamefont {Z.-X.}\
			\bibnamefont {Gong}}, \bibinfo {author} {\bibfnamefont {A.~V.}\ \bibnamefont
			{Gorshkov}},\ and\ \bibinfo {author} {\bibfnamefont {C.}~\bibnamefont
			{Monroe}},\ }\href {https://www.science.org/doi/10.1126/sciadv.1700672}
	{\bibfield  {journal} {\bibinfo  {journal} {Science advances}\ }\textbf
		{\bibinfo {volume} {3}},\ \bibinfo {pages} {e1700672} (\bibinfo {year}
		{2017})}\BibitemShut {NoStop}%
	\bibitem [{\citenamefont {Smith}\ \emph {et~al.}(2016)\citenamefont {Smith},
		\citenamefont {Lee}, \citenamefont {Richerme}, \citenamefont {Neyenhuis},
		\citenamefont {Hess}, \citenamefont {Hauke}, \citenamefont {Heyl},
		\citenamefont {Huse},\ and\ \citenamefont {Monroe}}]{smith2016many}%
	\BibitemOpen
	\bibfield  {author} {\bibinfo {author} {\bibfnamefont {J.}~\bibnamefont
			{Smith}}, \bibinfo {author} {\bibfnamefont {A.}~\bibnamefont {Lee}}, \bibinfo
		{author} {\bibfnamefont {P.}~\bibnamefont {Richerme}}, \bibinfo {author}
		{\bibfnamefont {B.}~\bibnamefont {Neyenhuis}}, \bibinfo {author}
		{\bibfnamefont {P.~W.}\ \bibnamefont {Hess}}, \bibinfo {author}
		{\bibfnamefont {P.}~\bibnamefont {Hauke}}, \bibinfo {author} {\bibfnamefont
			{M.}~\bibnamefont {Heyl}}, \bibinfo {author} {\bibfnamefont {D.~A.}\
			\bibnamefont {Huse}},\ and\ \bibinfo {author} {\bibfnamefont
			{C.}~\bibnamefont {Monroe}},\ }\href {https://doi.org/10.1038/nphys3783}
	{\bibfield  {journal} {\bibinfo  {journal} {Nature Physics}\ }\textbf
		{\bibinfo {volume} {12}},\ \bibinfo {pages} {907} (\bibinfo {year}
		{2016})}\BibitemShut {NoStop}%
	\bibitem [{\citenamefont {Choi}\ \emph {et~al.}(2016)\citenamefont {Choi},
		\citenamefont {Hild}, \citenamefont {Zeiher}, \citenamefont {Schau{\ss}},
		\citenamefont {Rubio-Abadal}, \citenamefont {Yefsah}, \citenamefont
		{Khemani}, \citenamefont {Huse}, \citenamefont {Bloch},\ and\ \citenamefont
		{Gross}}]{choi2016exploring}%
	\BibitemOpen
	\bibfield  {author} {\bibinfo {author} {\bibfnamefont {J.-Y.}\ \bibnamefont
			{Choi}}, \bibinfo {author} {\bibfnamefont {S.}~\bibnamefont {Hild}}, \bibinfo
		{author} {\bibfnamefont {J.}~\bibnamefont {Zeiher}}, \bibinfo {author}
		{\bibfnamefont {P.}~\bibnamefont {Schau{\ss}}}, \bibinfo {author}
		{\bibfnamefont {A.}~\bibnamefont {Rubio-Abadal}}, \bibinfo {author}
		{\bibfnamefont {T.}~\bibnamefont {Yefsah}}, \bibinfo {author} {\bibfnamefont
			{V.}~\bibnamefont {Khemani}}, \bibinfo {author} {\bibfnamefont {D.~A.}\
			\bibnamefont {Huse}}, \bibinfo {author} {\bibfnamefont {I.}~\bibnamefont
			{Bloch}},\ and\ \bibinfo {author} {\bibfnamefont {C.}~\bibnamefont {Gross}},\
	}\href {https://www.science.org/doi/10.1126/science.aaf8834} {\bibfield
		{journal} {\bibinfo  {journal} {Science}\ }\textbf {\bibinfo {volume}
			{352}},\ \bibinfo {pages} {1547} (\bibinfo {year} {2016})}\BibitemShut
	{NoStop}%
	\bibitem [{\citenamefont {Zhang}\ \emph
		{et~al.}(2017{\natexlab{a}})\citenamefont {Zhang}, \citenamefont {Hess},
		\citenamefont {Kyprianidis}, \citenamefont {Becker}, \citenamefont {Lee},
		\citenamefont {Smith}, \citenamefont {Pagano}, \citenamefont {Potirniche},
		\citenamefont {Potter}, \citenamefont {Vishwanath} \emph
		{et~al.}}]{zhang2017observation1}%
	\BibitemOpen
	\bibfield  {author} {\bibinfo {author} {\bibfnamefont {J.}~\bibnamefont
			{Zhang}}, \bibinfo {author} {\bibfnamefont {P.~W.}\ \bibnamefont {Hess}},
		\bibinfo {author} {\bibfnamefont {A.}~\bibnamefont {Kyprianidis}}, \bibinfo
		{author} {\bibfnamefont {P.}~\bibnamefont {Becker}}, \bibinfo {author}
		{\bibfnamefont {A.}~\bibnamefont {Lee}}, \bibinfo {author} {\bibfnamefont
			{J.}~\bibnamefont {Smith}}, \bibinfo {author} {\bibfnamefont
			{G.}~\bibnamefont {Pagano}}, \bibinfo {author} {\bibfnamefont {I.-D.}\
			\bibnamefont {Potirniche}}, \bibinfo {author} {\bibfnamefont {A.~C.}\
			\bibnamefont {Potter}}, \bibinfo {author} {\bibfnamefont {A.}~\bibnamefont
			{Vishwanath}}, \emph {et~al.},\ }\href {https://doi.org/10.1038/nature21413}
	{\bibfield  {journal} {\bibinfo  {journal} {Nature}\ }\textbf {\bibinfo
			{volume} {543}},\ \bibinfo {pages} {217} (\bibinfo {year}
		{2017}{\natexlab{a}})}\BibitemShut {NoStop}%
	\bibitem [{\citenamefont {Zhang}\ \emph
		{et~al.}(2017{\natexlab{b}})\citenamefont {Zhang}, \citenamefont {Pagano},
		\citenamefont {Hess}, \citenamefont {Kyprianidis}, \citenamefont {Becker},
		\citenamefont {Kaplan}, \citenamefont {Gorshkov}, \citenamefont {Gong},\ and\
		\citenamefont {Monroe}}]{zhang2017observation2}%
	\BibitemOpen
	\bibfield  {author} {\bibinfo {author} {\bibfnamefont {J.}~\bibnamefont
			{Zhang}}, \bibinfo {author} {\bibfnamefont {G.}~\bibnamefont {Pagano}},
		\bibinfo {author} {\bibfnamefont {P.~W.}\ \bibnamefont {Hess}}, \bibinfo
		{author} {\bibfnamefont {A.}~\bibnamefont {Kyprianidis}}, \bibinfo {author}
		{\bibfnamefont {P.}~\bibnamefont {Becker}}, \bibinfo {author} {\bibfnamefont
			{H.}~\bibnamefont {Kaplan}}, \bibinfo {author} {\bibfnamefont {A.~V.}\
			\bibnamefont {Gorshkov}}, \bibinfo {author} {\bibfnamefont {Z.-X.}\
			\bibnamefont {Gong}},\ and\ \bibinfo {author} {\bibfnamefont
			{C.}~\bibnamefont {Monroe}},\ }\href {https://doi.org/10.1038/nature24654}
	{\bibfield  {journal} {\bibinfo  {journal} {Nature}\ }\textbf {\bibinfo
			{volume} {551}},\ \bibinfo {pages} {601} (\bibinfo {year}
		{2017}{\natexlab{b}})}\BibitemShut {NoStop}%
	\bibitem [{\citenamefont {Jurcevic}\ \emph {et~al.}(2017)\citenamefont
		{Jurcevic}, \citenamefont {Shen}, \citenamefont {Hauke}, \citenamefont
		{Maier}, \citenamefont {Brydges}, \citenamefont {Hempel}, \citenamefont
		{Lanyon}, \citenamefont {Heyl}, \citenamefont {Blatt},\ and\ \citenamefont
		{Roos}}]{jurcevic2017direct}%
	\BibitemOpen
	\bibfield  {author} {\bibinfo {author} {\bibfnamefont {P.}~\bibnamefont
			{Jurcevic}}, \bibinfo {author} {\bibfnamefont {H.}~\bibnamefont {Shen}},
		\bibinfo {author} {\bibfnamefont {P.}~\bibnamefont {Hauke}}, \bibinfo
		{author} {\bibfnamefont {C.}~\bibnamefont {Maier}}, \bibinfo {author}
		{\bibfnamefont {T.}~\bibnamefont {Brydges}}, \bibinfo {author} {\bibfnamefont
			{C.}~\bibnamefont {Hempel}}, \bibinfo {author} {\bibfnamefont
			{B.}~\bibnamefont {Lanyon}}, \bibinfo {author} {\bibfnamefont
			{M.}~\bibnamefont {Heyl}}, \bibinfo {author} {\bibfnamefont {R.}~\bibnamefont
			{Blatt}},\ and\ \bibinfo {author} {\bibfnamefont {C.}~\bibnamefont {Roos}},\
	}\href {https://doi.org/10.1103/PhysRevLett.119.080501} {\bibfield  {journal}
		{\bibinfo  {journal} {Physical review letters}\ }\textbf {\bibinfo {volume}
			{119}},\ \bibinfo {pages} {080501} (\bibinfo {year} {2017})}\BibitemShut
	{NoStop}%
	\bibitem [{\citenamefont {Jotzu}\ \emph {et~al.}(2014)\citenamefont {Jotzu},
		\citenamefont {Messer}, \citenamefont {Desbuquois}, \citenamefont {Lebrat},
		\citenamefont {Uehlinger}, \citenamefont {Greif},\ and\ \citenamefont
		{Esslinger}}]{jotzu2014experimental}%
	\BibitemOpen
	\bibfield  {author} {\bibinfo {author} {\bibfnamefont {G.}~\bibnamefont
			{Jotzu}}, \bibinfo {author} {\bibfnamefont {M.}~\bibnamefont {Messer}},
		\bibinfo {author} {\bibfnamefont {R.}~\bibnamefont {Desbuquois}}, \bibinfo
		{author} {\bibfnamefont {M.}~\bibnamefont {Lebrat}}, \bibinfo {author}
		{\bibfnamefont {T.}~\bibnamefont {Uehlinger}}, \bibinfo {author}
		{\bibfnamefont {D.}~\bibnamefont {Greif}},\ and\ \bibinfo {author}
		{\bibfnamefont {T.}~\bibnamefont {Esslinger}},\ }\href
	{https://doi.org/10.1038/nature13915} {\bibfield  {journal} {\bibinfo
			{journal} {Nature}\ }\textbf {\bibinfo {volume} {515}},\ \bibinfo {pages}
		{237} (\bibinfo {year} {2014})}\BibitemShut {NoStop}%
	\bibitem [{\citenamefont {Wu}\ \emph {et~al.}(2016)\citenamefont {Wu},
		\citenamefont {Zhang}, \citenamefont {Sun}, \citenamefont {Xu}, \citenamefont
		{Wang}, \citenamefont {Ji}, \citenamefont {Deng}, \citenamefont {Chen},
		\citenamefont {Liu},\ and\ \citenamefont {Pan}}]{wu2016realization}%
	\BibitemOpen
	\bibfield  {author} {\bibinfo {author} {\bibfnamefont {Z.}~\bibnamefont
			{Wu}}, \bibinfo {author} {\bibfnamefont {L.}~\bibnamefont {Zhang}}, \bibinfo
		{author} {\bibfnamefont {W.}~\bibnamefont {Sun}}, \bibinfo {author}
		{\bibfnamefont {X.-T.}\ \bibnamefont {Xu}}, \bibinfo {author} {\bibfnamefont
			{B.-Z.}\ \bibnamefont {Wang}}, \bibinfo {author} {\bibfnamefont {S.-C.}\
			\bibnamefont {Ji}}, \bibinfo {author} {\bibfnamefont {Y.}~\bibnamefont
			{Deng}}, \bibinfo {author} {\bibfnamefont {S.}~\bibnamefont {Chen}}, \bibinfo
		{author} {\bibfnamefont {X.-J.}\ \bibnamefont {Liu}},\ and\ \bibinfo {author}
		{\bibfnamefont {J.-W.}\ \bibnamefont {Pan}},\ }\href
	{https://doi.org/10.1126/science.aaf6689} {\bibfield  {journal} {\bibinfo
			{journal} {Science}\ }\textbf {\bibinfo {volume} {354}},\ \bibinfo {pages}
		{83} (\bibinfo {year} {2016})}\BibitemShut {NoStop}%
	\bibitem [{\citenamefont {Lohse}\ \emph {et~al.}(2016)\citenamefont {Lohse},
		\citenamefont {Schweizer}, \citenamefont {Zilberberg}, \citenamefont
		{Aidelsburger},\ and\ \citenamefont {Bloch}}]{lohse2016thouless}%
	\BibitemOpen
	\bibfield  {author} {\bibinfo {author} {\bibfnamefont {M.}~\bibnamefont
			{Lohse}}, \bibinfo {author} {\bibfnamefont {C.}~\bibnamefont {Schweizer}},
		\bibinfo {author} {\bibfnamefont {O.}~\bibnamefont {Zilberberg}}, \bibinfo
		{author} {\bibfnamefont {M.}~\bibnamefont {Aidelsburger}},\ and\ \bibinfo
		{author} {\bibfnamefont {I.}~\bibnamefont {Bloch}},\ }\href
	{https://doi.org/10.1038/nphys3584} {\bibfield  {journal} {\bibinfo
			{journal} {Nature Physics}\ }\textbf {\bibinfo {volume} {12}},\ \bibinfo
		{pages} {350} (\bibinfo {year} {2016})}\BibitemShut {NoStop}%
	\bibitem [{\citenamefont {Tai}\ \emph {et~al.}(2017)\citenamefont {Tai},
		\citenamefont {Lukin}, \citenamefont {Rispoli}, \citenamefont {Schittko},
		\citenamefont {Menke}, \citenamefont {Borgnia}, \citenamefont {Preiss},
		\citenamefont {Grusdt}, \citenamefont {Kaufman},\ and\ \citenamefont
		{Greiner}}]{tai2017microscopy}%
	\BibitemOpen
	\bibfield  {author} {\bibinfo {author} {\bibfnamefont {M.~E.}\ \bibnamefont
			{Tai}}, \bibinfo {author} {\bibfnamefont {A.}~\bibnamefont {Lukin}}, \bibinfo
		{author} {\bibfnamefont {M.}~\bibnamefont {Rispoli}}, \bibinfo {author}
		{\bibfnamefont {R.}~\bibnamefont {Schittko}}, \bibinfo {author}
		{\bibfnamefont {T.}~\bibnamefont {Menke}}, \bibinfo {author} {\bibfnamefont
			{D.}~\bibnamefont {Borgnia}}, \bibinfo {author} {\bibfnamefont {P.~M.}\
			\bibnamefont {Preiss}}, \bibinfo {author} {\bibfnamefont {F.}~\bibnamefont
			{Grusdt}}, \bibinfo {author} {\bibfnamefont {A.~M.}\ \bibnamefont
			{Kaufman}},\ and\ \bibinfo {author} {\bibfnamefont {M.}~\bibnamefont
			{Greiner}},\ }\href {https://doi.org/10.1038/nature22811} {\bibfield
		{journal} {\bibinfo  {journal} {Nature}\ }\textbf {\bibinfo {volume} {546}},\
		\bibinfo {pages} {519} (\bibinfo {year} {2017})}\BibitemShut {NoStop}%
	\bibitem [{\citenamefont {Kennedy}\ \emph {et~al.}(2015)\citenamefont
		{Kennedy}, \citenamefont {Burton}, \citenamefont {Chung},\ and\ \citenamefont
		{Ketterle}}]{kennedy2015observation}%
	\BibitemOpen
	\bibfield  {author} {\bibinfo {author} {\bibfnamefont {C.~J.}\ \bibnamefont
			{Kennedy}}, \bibinfo {author} {\bibfnamefont {W.~C.}\ \bibnamefont {Burton}},
		\bibinfo {author} {\bibfnamefont {W.~C.}\ \bibnamefont {Chung}},\ and\
		\bibinfo {author} {\bibfnamefont {W.}~\bibnamefont {Ketterle}},\ }\href
	{https://doi.org/10.1038/nphys3421} {\bibfield  {journal} {\bibinfo
			{journal} {Nature Physics}\ }\textbf {\bibinfo {volume} {11}},\ \bibinfo
		{pages} {859} (\bibinfo {year} {2015})}\BibitemShut {NoStop}%
	\bibitem [{\citenamefont {Fl{\"a}schner}\ \emph {et~al.}(2018)\citenamefont
		{Fl{\"a}schner}, \citenamefont {Vogel}, \citenamefont {Tarnowski},
		\citenamefont {Rem}, \citenamefont {L{\"u}hmann}, \citenamefont {Heyl},
		\citenamefont {Budich}, \citenamefont {Mathey}, \citenamefont {Sengstock},\
		and\ \citenamefont {Weitenberg}}]{flaschner2018observation}%
	\BibitemOpen
	\bibfield  {author} {\bibinfo {author} {\bibfnamefont {N.}~\bibnamefont
			{Fl{\"a}schner}}, \bibinfo {author} {\bibfnamefont {D.}~\bibnamefont
			{Vogel}}, \bibinfo {author} {\bibfnamefont {M.}~\bibnamefont {Tarnowski}},
		\bibinfo {author} {\bibfnamefont {B.}~\bibnamefont {Rem}}, \bibinfo {author}
		{\bibfnamefont {D.-S.}\ \bibnamefont {L{\"u}hmann}}, \bibinfo {author}
		{\bibfnamefont {M.}~\bibnamefont {Heyl}}, \bibinfo {author} {\bibfnamefont
			{J.}~\bibnamefont {Budich}}, \bibinfo {author} {\bibfnamefont
			{L.}~\bibnamefont {Mathey}}, \bibinfo {author} {\bibfnamefont
			{K.}~\bibnamefont {Sengstock}},\ and\ \bibinfo {author} {\bibfnamefont
			{C.}~\bibnamefont {Weitenberg}},\ }\href
	{https://doi.org/10.1038/s41567-017-0013-8} {\bibfield  {journal} {\bibinfo
			{journal} {Nature Physics}\ }\textbf {\bibinfo {volume} {14}},\ \bibinfo
		{pages} {265} (\bibinfo {year} {2018})}\BibitemShut {NoStop}%
	\bibitem [{\citenamefont {Heyl}\ \emph {et~al.}(2013)\citenamefont {Heyl},
		\citenamefont {Polkovnikov},\ and\ \citenamefont
		{Kehrein}}]{heyl2013dynamical}%
	\BibitemOpen
	\bibfield  {author} {\bibinfo {author} {\bibfnamefont {M.}~\bibnamefont
			{Heyl}}, \bibinfo {author} {\bibfnamefont {A.}~\bibnamefont {Polkovnikov}},\
		and\ \bibinfo {author} {\bibfnamefont {S.}~\bibnamefont {Kehrein}},\ }\href
	{https://doi.org/10.1103/PhysRevLett.110.135704} {\bibfield  {journal}
		{\bibinfo  {journal} {Physical review letters}\ }\textbf {\bibinfo {volume}
			{110}},\ \bibinfo {pages} {135704} (\bibinfo {year} {2013})}\BibitemShut
	{NoStop}%
	\bibitem [{\citenamefont {Heyl}(2018)}]{heyl2018dynamical}%
	\BibitemOpen
	\bibfield  {author} {\bibinfo {author} {\bibfnamefont {M.}~\bibnamefont
			{Heyl}},\ }\href {https://dx.doi.org/10.1088/1361-6633/aaaf9a} {\bibfield
		{journal} {\bibinfo  {journal} {Reports on Progress in Physics}\ }\textbf
		{\bibinfo {volume} {81}},\ \bibinfo {pages} {054001} (\bibinfo {year}
		{2018})}\BibitemShut {NoStop}%
	\bibitem [{\citenamefont {Wang}\ \emph {et~al.}(2017)\citenamefont {Wang},
		\citenamefont {Zhang}, \citenamefont {Chen}, \citenamefont {Yu},\ and\
		\citenamefont {Zhai}}]{wang2017scheme}%
	\BibitemOpen
	\bibfield  {author} {\bibinfo {author} {\bibfnamefont {C.}~\bibnamefont
			{Wang}}, \bibinfo {author} {\bibfnamefont {P.}~\bibnamefont {Zhang}},
		\bibinfo {author} {\bibfnamefont {X.}~\bibnamefont {Chen}}, \bibinfo {author}
		{\bibfnamefont {J.}~\bibnamefont {Yu}},\ and\ \bibinfo {author}
		{\bibfnamefont {H.}~\bibnamefont {Zhai}},\ }\href
	{https://doi.org/10.1103/PhysRevLett.118.185701} {\bibfield  {journal}
		{\bibinfo  {journal} {Physical review letters}\ }\textbf {\bibinfo {volume}
			{118}},\ \bibinfo {pages} {185701} (\bibinfo {year} {2017})}\BibitemShut
	{NoStop}%
	\bibitem [{\citenamefont {Heyl}(2015)}]{heyl2015scaling}%
	\BibitemOpen
	\bibfield  {author} {\bibinfo {author} {\bibfnamefont {M.}~\bibnamefont
			{Heyl}},\ }\href {https://doi.org/10.1103/PhysRevLett.115.140602} {\bibfield
		{journal} {\bibinfo  {journal} {Physical Review Letters}\ }\textbf {\bibinfo
			{volume} {115}},\ \bibinfo {pages} {140602} (\bibinfo {year}
		{2015})}\BibitemShut {NoStop}%
	\bibitem [{\citenamefont {Shi}\ \emph {et~al.}(2022)\citenamefont {Shi},
		\citenamefont {Zhang},\ and\ \citenamefont {Song}}]{shi2022dynamic}%
	\BibitemOpen
	\bibfield  {author} {\bibinfo {author} {\bibfnamefont {Y.}~\bibnamefont
			{Shi}}, \bibinfo {author} {\bibfnamefont {K.}~\bibnamefont {Zhang}},\ and\
		\bibinfo {author} {\bibfnamefont {Z.}~\bibnamefont {Song}},\ }\href
	{https://doi.org/10.1103/RevModPhys.93.025001} {\bibfield  {journal}
		{\bibinfo  {journal} {Physical Review B}\ }\textbf {\bibinfo {volume}
			{106}},\ \bibinfo {pages} {184505} (\bibinfo {year} {2022})}\BibitemShut
	{NoStop}%
	\bibitem [{\citenamefont {Altland}\ and\ \citenamefont
		{Zirnbauer}(1997)}]{altland1997nonstandard}%
	\BibitemOpen
	\bibfield  {author} {\bibinfo {author} {\bibfnamefont {A.}~\bibnamefont
			{Altland}}\ and\ \bibinfo {author} {\bibfnamefont {M.~R.}\ \bibnamefont
			{Zirnbauer}},\ }\href
	{https://journals.aps.org/prb/abstract/10.1103/PhysRevB.55.1142} {\bibfield
		{journal} {\bibinfo  {journal} {Physical Review B}\ }\textbf {\bibinfo
			{volume} {55}},\ \bibinfo {pages} {1142} (\bibinfo {year}
		{1997})}\BibitemShut {NoStop}%
	\bibitem [{\citenamefont {Kitaev}(2001)}]{kitaev2001unpaired}%
	\BibitemOpen
	\bibfield  {author} {\bibinfo {author} {\bibfnamefont {A.~Y.}\ \bibnamefont
			{Kitaev}},\ }\href
	{https://iopscience.iop.org/article/10.1070/1063-7869/44/10S/S29} {\bibfield
		{journal} {\bibinfo  {journal} {Physics-uspekhi}\ }\textbf {\bibinfo {volume}
			{44}},\ \bibinfo {pages} {131} (\bibinfo {year} {2001})}\BibitemShut
	{NoStop}%
	\bibitem [{\citenamefont {Soori}(2024)}]{soori2024majorana}%
	\BibitemOpen
	\bibfield  {author} {\bibinfo {author} {\bibfnamefont {A.}~\bibnamefont
			{Soori}},\ }\href {https://doi.org/10.48550/arXiv.2403.02266} {\bibfield
		{journal} {\bibinfo  {journal} {arXiv preprint arXiv:2403.02266}\ } (\bibinfo
		{year} {2024})}\BibitemShut {NoStop}%
	\bibitem [{\citenamefont {Decker}\ and\ \citenamefont
		{Karrasch}(2024)}]{decker2024density}%
	\BibitemOpen
	\bibfield  {author} {\bibinfo {author} {\bibfnamefont {K.}~\bibnamefont
			{Decker}}\ and\ \bibinfo {author} {\bibfnamefont {C.}~\bibnamefont
			{Karrasch}},\ }\href {https://doi.org/10.48550/arXiv.2402.12897} {\bibfield
		{journal} {\bibinfo  {journal} {arXiv preprint arXiv:2402.12897}\ } (\bibinfo
		{year} {2024})}\BibitemShut {NoStop}%
	\bibitem [{\citenamefont {Malard}\ and\ \citenamefont
		{Brand{\~a}o}(2024)}]{malarddetecting}%
	\BibitemOpen
	\bibfield  {author} {\bibinfo {author} {\bibfnamefont {M.}~\bibnamefont
			{Malard}}\ and\ \bibinfo {author} {\bibfnamefont {D.~S.}\ \bibnamefont
			{Brand{\~a}o}},\ }\href {https://doi.org/10.48550/arXiv.2403.01588 Focus to
		learn more} {\bibfield  {journal} {\bibinfo  {journal} {arXiv preprint
				arXiv:2403.01588}\ } (\bibinfo {year} {2024})}\BibitemShut {NoStop}%
	\bibitem [{\citenamefont {Silva}\ \emph {et~al.}(2024)\citenamefont {Silva},
		\citenamefont {Ribeiro}, \citenamefont {Caldas},\ and\ \citenamefont
		{Continentino}}]{silva2024hybridization}%
	\BibitemOpen
	\bibfield  {author} {\bibinfo {author} {\bibfnamefont {E.}~\bibnamefont
			{Silva}}, \bibinfo {author} {\bibfnamefont {R.~B.}\ \bibnamefont {Ribeiro}},
		\bibinfo {author} {\bibfnamefont {H.}~\bibnamefont {Caldas}},\ and\ \bibinfo
		{author} {\bibfnamefont {M.~A.}\ \bibnamefont {Continentino}},\ }\href
	{https://doi.org/10.1103/PhysRevB.109.134503} {\bibfield  {journal} {\bibinfo
			{journal} {Physical Review B}\ }\textbf {\bibinfo {volume} {109}},\ \bibinfo
		{pages} {134503} (\bibinfo {year} {2024})}\BibitemShut {NoStop}%
	\bibitem [{\citenamefont {Starchl}\ and\ \citenamefont
		{Sieberer}(2022)}]{starchl2022relaxation}%
	\BibitemOpen
	\bibfield  {author} {\bibinfo {author} {\bibfnamefont {E.}~\bibnamefont
			{Starchl}}\ and\ \bibinfo {author} {\bibfnamefont {L.~M.}\ \bibnamefont
			{Sieberer}},\ }\href {https://doi.org/10.1103/PhysRevLett.129.220602}
	{\bibfield  {journal} {\bibinfo  {journal} {Physical Review Letters}\
		}\textbf {\bibinfo {volume} {129}},\ \bibinfo {pages} {220602} (\bibinfo
		{year} {2022})}\BibitemShut {NoStop}%
	\bibitem [{\citenamefont {Shi}\ \emph {et~al.}(2023)\citenamefont {Shi},
		\citenamefont {Zhang},\ and\ \citenamefont {Song}}]{shi2023emerging}%
	\BibitemOpen
	\bibfield  {author} {\bibinfo {author} {\bibfnamefont {Y.}~\bibnamefont
			{Shi}}, \bibinfo {author} {\bibfnamefont {X.}~\bibnamefont {Zhang}},\ and\
		\bibinfo {author} {\bibfnamefont {Z.}~\bibnamefont {Song}},\ }\href
	{https://doi.org/10.48550/arXiv.2311.08056} {\bibfield  {journal} {\bibinfo
			{journal} {arXiv preprint arXiv:2311.08056}\ } (\bibinfo {year}
		{2023})}\BibitemShut {NoStop}%
	\bibitem [{\citenamefont {Kosior}\ and\ \citenamefont
		{Heyl}(2024)}]{kosior2024vortex}%
	\BibitemOpen
	\bibfield  {author} {\bibinfo {author} {\bibfnamefont {A.}~\bibnamefont
			{Kosior}}\ and\ \bibinfo {author} {\bibfnamefont {M.}~\bibnamefont {Heyl}},\
	}\href {https://doi.org/10.1103/PhysRevB.109.L140303} {\bibfield  {journal}
		{\bibinfo  {journal} {Physical Review B}\ }\textbf {\bibinfo {volume}
			{109}},\ \bibinfo {pages} {L140303} (\bibinfo {year} {2024})}\BibitemShut
	{NoStop}%
	\bibitem [{\citenamefont {Yang}\ and\ \citenamefont
		{Song}(2020)}]{yang2020biot}%
	\BibitemOpen
	\bibfield  {author} {\bibinfo {author} {\bibfnamefont {X.}~\bibnamefont
			{Yang}}\ and\ \bibinfo {author} {\bibfnamefont {Z.}~\bibnamefont {Song}},\
	}\href {https://doi.org/10.1103/PhysRevB.102.195112} {\bibfield  {journal}
		{\bibinfo  {journal} {Physical Review B}\ }\textbf {\bibinfo {volume}
			{102}},\ \bibinfo {pages} {195112} (\bibinfo {year} {2020})}\BibitemShut
	{NoStop}%
	\bibitem [{\citenamefont {Qi}\ \emph {et~al.}(2006)\citenamefont {Qi},
		\citenamefont {Wu},\ and\ \citenamefont {Zhang}}]{qi2006topological}%
	\BibitemOpen
	\bibfield  {author} {\bibinfo {author} {\bibfnamefont {X.-L.}\ \bibnamefont
			{Qi}}, \bibinfo {author} {\bibfnamefont {Y.-S.}\ \bibnamefont {Wu}},\ and\
		\bibinfo {author} {\bibfnamefont {S.-C.}\ \bibnamefont {Zhang}},\ }\href
	{https://doi.org/10.1103/PhysRevB.74.085308} {\bibfield  {journal} {\bibinfo
			{journal} {Physical Review B—Condensed Matter and Materials Physics}\
		}\textbf {\bibinfo {volume} {74}},\ \bibinfo {pages} {085308} (\bibinfo
		{year} {2006})}\BibitemShut {NoStop}%
	\bibitem [{\citenamefont {Bernevig}\ \emph {et~al.}(2006)\citenamefont
		{Bernevig}, \citenamefont {Hughes},\ and\ \citenamefont
		{Zhang}}]{bernevig2006quantum}%
	\BibitemOpen
	\bibfield  {author} {\bibinfo {author} {\bibfnamefont {B.~A.}\ \bibnamefont
			{Bernevig}}, \bibinfo {author} {\bibfnamefont {T.~L.}\ \bibnamefont
			{Hughes}},\ and\ \bibinfo {author} {\bibfnamefont {S.-C.}\ \bibnamefont
			{Zhang}},\ }\href {10.1126/science.1133734} {\bibfield  {journal} {\bibinfo
			{journal} {science}\ }\textbf {\bibinfo {volume} {314}},\ \bibinfo {pages}
		{1757} (\bibinfo {year} {2006})}\BibitemShut {NoStop}%
	\bibitem [{\citenamefont {Qi}\ and\ \citenamefont
		{Zhang}(2011)}]{qi2011topological}%
	\BibitemOpen
	\bibfield  {author} {\bibinfo {author} {\bibfnamefont {X.-L.}\ \bibnamefont
			{Qi}}\ and\ \bibinfo {author} {\bibfnamefont {S.-C.}\ \bibnamefont {Zhang}},\
	}\href {https://doi.org/10.1103/RevModPhys.83.1057} {\bibfield  {journal}
		{\bibinfo  {journal} {Reviews of modern physics}\ }\textbf {\bibinfo {volume}
			{83}},\ \bibinfo {pages} {1057} (\bibinfo {year} {2011})}\BibitemShut
	{NoStop}%
	\bibitem [{\citenamefont {Cho}\ and\ \citenamefont
		{Moore}(2011)}]{cho2011quantum}%
	\BibitemOpen
	\bibfield  {author} {\bibinfo {author} {\bibfnamefont {G.~Y.}\ \bibnamefont
			{Cho}}\ and\ \bibinfo {author} {\bibfnamefont {J.~E.}\ \bibnamefont
			{Moore}},\ }\href {https://doi.org/10.1103/PhysRevB.84.165101} {\bibfield
		{journal} {\bibinfo  {journal} {Physical Review B—Condensed Matter and
				Materials Physics}\ }\textbf {\bibinfo {volume} {84}},\ \bibinfo {pages}
		{165101} (\bibinfo {year} {2011})}\BibitemShut {NoStop}%
	\bibitem [{\citenamefont {Yi}\ \emph {et~al.}(2019)\citenamefont {Yi},
		\citenamefont {Zhang}, \citenamefont {Zhang}, \citenamefont {Jiao},
		\citenamefont {Cheng}, \citenamefont {Wang}, \citenamefont {Xu},
		\citenamefont {Sun}, \citenamefont {Liu}, \citenamefont {Chen} \emph
		{et~al.}}]{yi2019observing}%
	\BibitemOpen
	\bibfield  {author} {\bibinfo {author} {\bibfnamefont {C.-R.}\ \bibnamefont
			{Yi}}, \bibinfo {author} {\bibfnamefont {L.}~\bibnamefont {Zhang}}, \bibinfo
		{author} {\bibfnamefont {L.}~\bibnamefont {Zhang}}, \bibinfo {author}
		{\bibfnamefont {R.-H.}\ \bibnamefont {Jiao}}, \bibinfo {author}
		{\bibfnamefont {X.-C.}\ \bibnamefont {Cheng}}, \bibinfo {author}
		{\bibfnamefont {Z.-Y.}\ \bibnamefont {Wang}}, \bibinfo {author}
		{\bibfnamefont {X.-T.}\ \bibnamefont {Xu}}, \bibinfo {author} {\bibfnamefont
			{W.}~\bibnamefont {Sun}}, \bibinfo {author} {\bibfnamefont {X.-J.}\
			\bibnamefont {Liu}}, \bibinfo {author} {\bibfnamefont {S.}~\bibnamefont
			{Chen}}, \emph {et~al.},\ }\href
	{https://doi.org/10.1103/PhysRevLett.123.190603} {\bibfield  {journal}
		{\bibinfo  {journal} {Physical review letters}\ }\textbf {\bibinfo {volume}
			{123}},\ \bibinfo {pages} {190603} (\bibinfo {year} {2019})}\BibitemShut
	{NoStop}%
	\bibitem [{\citenamefont {Ma}\ \emph {et~al.}(2023)\citenamefont {Ma},
		\citenamefont {Li}, \citenamefont {Wang}, \citenamefont {Zhao}, \citenamefont
		{Xiong},\ and\ \citenamefont {Sun}}]{ma2023orthogonal}%
	\BibitemOpen
	\bibfield  {author} {\bibinfo {author} {\bibfnamefont {Y.}~\bibnamefont
			{Ma}}, \bibinfo {author} {\bibfnamefont {X.}~\bibnamefont {Li}}, \bibinfo
		{author} {\bibfnamefont {Y.}~\bibnamefont {Wang}}, \bibinfo {author}
		{\bibfnamefont {S.}~\bibnamefont {Zhao}}, \bibinfo {author} {\bibfnamefont
			{G.}~\bibnamefont {Xiong}},\ and\ \bibinfo {author} {\bibfnamefont
			{T.}~\bibnamefont {Sun}},\ }\href
	{https://doi.org/10.1103/PhysRevB.108.075128} {\bibfield  {journal} {\bibinfo
			{journal} {Physical Review B}\ }\textbf {\bibinfo {volume} {108}},\ \bibinfo
		{pages} {075128} (\bibinfo {year} {2023})}\BibitemShut {NoStop}%
	\bibitem [{\citenamefont {Mittal}\ \emph {et~al.}(2018)\citenamefont {Mittal},
		\citenamefont {Goldschmidt},\ and\ \citenamefont
		{Hafezi}}]{mittal2018topological}%
	\BibitemOpen
	\bibfield  {author} {\bibinfo {author} {\bibfnamefont {S.}~\bibnamefont
			{Mittal}}, \bibinfo {author} {\bibfnamefont {E.~A.}\ \bibnamefont
			{Goldschmidt}},\ and\ \bibinfo {author} {\bibfnamefont {M.}~\bibnamefont
			{Hafezi}},\ }\href {https://doi.org/10.1038/s41586-018-0478-3} {\bibfield
		{journal} {\bibinfo  {journal} {Nature}\ }\textbf {\bibinfo {volume} {561}},\
		\bibinfo {pages} {502} (\bibinfo {year} {2018})}\BibitemShut {NoStop}%
	\bibitem [{\citenamefont {Hofmann}\ \emph {et~al.}(2019)\citenamefont
		{Hofmann}, \citenamefont {Helbig}, \citenamefont {Lee}, \citenamefont
		{Greiter},\ and\ \citenamefont {Thomale}}]{hofmann2019chiral}%
	\BibitemOpen
	\bibfield  {author} {\bibinfo {author} {\bibfnamefont {T.}~\bibnamefont
			{Hofmann}}, \bibinfo {author} {\bibfnamefont {T.}~\bibnamefont {Helbig}},
		\bibinfo {author} {\bibfnamefont {C.~H.}\ \bibnamefont {Lee}}, \bibinfo
		{author} {\bibfnamefont {M.}~\bibnamefont {Greiter}},\ and\ \bibinfo {author}
		{\bibfnamefont {R.}~\bibnamefont {Thomale}},\ }\href
	{https://doi.org/10.1103/PhysRevLett.122.247702} {\bibfield  {journal}
		{\bibinfo  {journal} {Physical review letters}\ }\textbf {\bibinfo {volume}
			{122}},\ \bibinfo {pages} {247702} (\bibinfo {year} {2019})}\BibitemShut
	{NoStop}%
\end{thebibliography}
\end{document}